\documentclass[prd,aps,showpacs,12pt,
nofootinbib,%
tightenlines,
superscriptaddress,
eqsecnum%
]{revtex4-1}

\pdfoutput=1 

\RequirePackage{color}
\RequirePackage[colorlinks=true
,urlcolor=blue
,anchorcolor=blue
,citecolor=blue
,filecolor=blue
,linkcolor=blue
,menucolor=blue
,linktocpage=true
,pdfproducer=medialab
,pdfa=true
]{hyperref}

\usepackage{graphicx}
\usepackage{amsmath}

\usepackage[T1]{fontenc} 

\usepackage{soul}
\sethlcolor{yellow}

\newcommand{\be}{\begin{equation}}
\newcommand{\ee}{\end{equation}}
\newcommand{\bea}{\begin{eqnarray}}
\newcommand{\eea}{\end{eqnarray}}

\newcommand{\zm}{z_0}

\begin{document}



\title{\boldmath Hadronic vacuum polarization contribution to \\ the muon $g-2$ in holographic QCD}


\author{Josef Leutgeb}
 \email{josef.leutgeb@tuwien.ac.at}
\author{Anton Rebhan}
 \email{anton.rebhan@tuwien.ac.at}
 \affiliation{Institut für Theoretische Physik, Technische Universität Wien, \\ Wiedner Hauptstrasse 8-10, A-1040 Vienna, Austria}
 
\author{Michael Stadlbauer}
 \email{michael.stadlbauer@tum.de}
 
\affiliation{Institut für Theoretische Physik, Technische Universität Wien, \\ Wiedner Hauptstrasse 8-10, A-1040 Vienna, Austria}
\affiliation{Technische Universität München, Physik-Department, \\ James-Franck-Strasse 1, 85748 Garching, Germany}
\affiliation{Max Planck Institute for Physics, Föhringer Ring 6, 80805 München, Germany}


\date{\today}

\begin{abstract}
We evaluate the leading-order hadronic vacuum polarization contribution to the anomalous magnetic moment of the muon with two light flavors in minimal hard-wall and soft-wall holographic QCD models, as well as in simple generalizations thereof, and compare with the
rather precise results available from dispersive and lattice approaches.
While holographic QCD cannot be expected to shed light on the existing small
discrepancies between the latter, this comparison in turn provides useful
information on the holographic models, which have been used to
evaluate hadronic light-by-light contributions where errors in data-driven
and lattice approaches are more sizable. In particular, in the hard-wall model that has recently been used
to implement the Melnikov-Vainshtein short-distance constraint on hadronic light-by-light contributions,
a matching of the hadronic vacuum
polarization to the data-driven approach points to the same correction of parameters that has been proposed recently in order to
account for next-to-leading order effects.

\end{abstract}


\maketitle

\tableofcontents 

\section{Introduction} \label{sec:introduction}

Currently there is a $4.2\sigma$ discrepancy between the Standard Model prediction
for
the anomalous magnetic moment of the muon $a_\mu=(g-2)_\mu/2$
as assembled in the White Paper (WP) \cite{Aoyama:2020ynm} and
the new experimental result, obtained by combining the BNL and the
Fermilab E989 values, involving an experimental error of $41\times 10^{-11}$,
which is expected to be reduced further by additional data taking and future experiments.
The uncertainty in the Standard Model prediction has a similar magnitude; it is
almost entirely due to
hadronic contributions from hadronic vacuum polarization (HVP)
and from hadronic light-by-light scattering (HLbL), which according to \cite{Aoyama:2020ynm}
amount to
\bea\label{WPHVP}
    a_\mu^{\text{HVP,WP}}&=&(6845 \pm 40)\times 10^{-11},\\
    a_\mu^{\text{HLbL,WP}}&=&(92 \pm 19)\times 10^{-11}.
\eea
The data-driven computation of HVP thus claims 
an accuracy of 0.6\%, whereas the much smaller HLbL contribution
has about 20\% uncertainty. To match the experimental progress, improvements
in the theoretical predictions for both contributions are called for.

However, the data-driven HVP result has recently been questioned by
a direct lattice QCD calculation \cite{Borsanyi:2020mff} by the BMW collaboration which claims a
similar error of $0.8\%$
\be\label{BMWHVP}
    a_\mu^{\text{HVP,BMW}}=(7075 \pm 55)\times 10^{-11}
\ee
but deviating from (\ref{WPHVP}) by about 3\% or $2.1\sigma$.
Taken at face value this would reduce the discrepancy between experiment and theory
in the case of $a_\mu$ to about $1.5\sigma$, while it may give rise to
tensions with electroweak precision fits of the hadronic contribution to the running of the electromagnetic coupling \cite{Passera:2008jk,Crivellin:2020zul,Keshavarzi:2020bfy}.

Once this critical issue has been resolved, it will also be crucial to
reduce the theoretical uncertainty in the HLbL contribution. In the latter,
an important question has been the implementation of certain
short-distance constraints in hadronic models \cite{Melnikov:2003xd,Ludtke:2020moa,Colangelo:2021nkr}, where
recently holographic QCD has helped to shed light on the role of
axial-vector mesons \cite{Leutgeb:2019gbz,Cappiello:2019hwh,Leutgeb:2021mpu,Leutgeb:2021bpo}.
Holographic QCD also makes interesting quantitative predictions given the large spread
of results in other hadronic models, which have led to a 100\% uncertainty
for the estimated contribution of axial-vector mesons in the WP.
Being an approach which is based on the large color number $N_c$
limit, it cannot be expected to help with the percent-level discrepancies
in the highly constrained HVP contribution. However, given that it typically achieves
an accuracy of 10-20\%, it can provide potentially useful information
in the case of HLbL. Investigating the performance in the case of HVP allows us
to test the holographic QCD models with regard to their ability to describe
photon-hadron interactions quantitatively.

In this paper we 
evaluate the leading-order HVP (LO-HVP) contribution
of the minimal bottom-up holographic QCD models that have been employed
in the study of the HLbL contribution, as well as simple generalizations
thereof, and compare with relevant results obtained
within the data-driven approach, in particular for the contributions
of the lightest quark flavors, revisiting and extending 
the study of Hong, Kim, and Matsuzaki \cite{Hong:2009jv}.


As shown in \cite{Blum:2002ii,Aubin:2006xv}, the leading order HVP contributions to the muon $g-2$ is related to the hadronic vacuum polarization function through
\begin{equation} \label{equ:amu_hvp}
a_{\mu}^{\text{LO-HVP}}=4\pi^2\left(\frac{\alpha}{\pi}\right)^{2} \int_{0}^{\infty} d Q^{2} f(Q^{2}) \Pi_{\text{em}}^{\mathrm{had}}(Q^{2}),
\end{equation}
where $Q^2=-q^2$ is the Euclidean momentum squared and
\begin{equation}
f(Q^2) = \frac{m^{2}_\mu Q^{2} Z^{3}\left(1-Q^{2} Z\right)}{1+m^{2}_\mu Q^{2} Z^{2}}, \quad Z=-\frac{Q^{2}-\sqrt{Q^{4}+4 m^{2}_\mu Q^{2}}}{2 m^{2}_\mu Q^{2}}.
\end{equation}
The hadronic vacuum polarization function needs to be renormalized
such that $\Pi_{\text{em}}^{\mathrm{had}}(0)=0$. It is given
by the vector-current correlator, defined by
\begin{equation}
i \int d^{4} x e^{i q x}\langle 0|T \{ J_{V}^{a \mu}(x) J_{V}^{b \nu}(0) \} | 0\rangle=\delta^{a b}\left(q^{2} \eta^{\mu \nu}-q^{\mu} q^{\nu}\right) \Pi_{V}\left(-q^{2}\right)
\end{equation}
in the flavor-symmetric case,
via
\begin{equation}
\Pi_{\text{em}}^{\mathrm{had}}\left(-q^{2}\right)=2 \operatorname{Tr} Q_{\text{em}}^{2} \Pi_{V}\left(-q^{2}\right),
\end{equation}
where $Q_{\text{em}}=\operatorname{diag}(\frac23,-\frac13,\ldots)$ is the quark charge matrix.

As we shall show, the holographic results can deviate by up to 50\% from
the data-driven result, even after accounting for the fact that the holographic
QCD results, being essentially a large-$N_c$ approximation, can only
account for a subset of multi-hadron intermediate states. 
However, the simplest HW model that can simultaneously
fit the rho meson mass and the pion decay constant as well as the
leading-order short-distance behavior of the vector correlator
is performing quite reasonably. Interestingly, the amount of correction expected
from next-to-leading order effects in the large-momentum domain, where perturbative
corrections to the asymptotic behavior proportional to $\alpha_s/\pi$ should
play a role, turn out to be consistent with the corrections proposed
recently by two of us in the case of the HLbL contribution \cite{Leutgeb:2021mpu}.

In the next section, we shall review the minimal holographic QCD models
included in our study to the extent necessary for evaluating the
LO-HVP contribution to $a_\mu$ in Sect.~\ref{sec:numerics}. Sect.~\ref{sec:conclusion}
summarizes our conclusions.


\section{Minimal holographic QCD models} 

In this work we shall limit ourselves to holographic QCD models with a minimal
set of adjustable parameters with anti-de Sitter background geometry and simple
generalizations thereof.

\subsection{Hard-wall models} \label{sec:HW_model}

In hard-wall (HW) AdS/QCD models, a five-dimensional anti-de Sitter (AdS) background
geometry is chosen. 
In terms of a holographic radial coordinate $z$ where the conformal
boundary is at $z=0$, the line element is given by (using a mostly-minus metric convention)
\begin{equation}\label{ds2AdS}
d s^{2}=g_{M N} d x^{M} d x^{N}=\frac{1}{z^{2}}\left(\eta_{\mu \nu} d x^{\mu} d x^{\nu}-d z^{2}\right), \quad 0<z \leq \zm.
\end{equation}
Conformal invariance is broken by a hard cutoff at $z=\zm$, where suitable
boundary conditions for bulk fields dual to the quantum operators of the
four-dimensional (large-$N_c$) gauge theory are imposed.

The fields dual to left and right quark bilinears $\bar\psi \gamma^\mu T^a P_{L,R}\psi$
are five-dimensional $U(N_f)_{L,R}$ gauge fields $A^{(L,R)}$, where chiral symmetry
breaking can be implemented 
through spontaneous symmetry breaking by a bifundamental scalar $X$ \cite{Erlich:2005qh,DaRold:2005mxj} or
through different boundary conditions \cite{Hirn:2005nr} on vector and
axial-vector fields, $V=\frac12[A^{(L)}+A^{(R)}]$ and $A=\frac12[A^{(L)}-A^{(R)}]$, 
or both \cite{Domenech:2010aq}.

The five-dimensional action of models with a bifundamental scalar $X$ 
introduced first by Erlich et al.~\cite{Erlich:2005qh} (termed HW1 in the following) reads
\begin{equation} \label{equ:action_HW_model}
S=\int d^{5} x \sqrt{g} \operatorname{Tr}\left\{|D X|^{2}+3|X|^{2}-\frac{1}{4 g_{5}^{2}}\left(F_{L}^{2}+F_{R}^{2}\right)\right\},
\end{equation}
whereas the model of Hirn and Sanz (termed HW2) has only the Yang-Mills part.

In both cases, the field equations for transverse vector fields $\partial_\mu V^\mu=0$ 
are given by
\begin{equation} \label{equ:EOM_hw_model}
\partial_{z}\left(\frac{1}{z} \partial_{z} V_{\mu}^{a}(q, z)\right)+\frac{q^{2}}{z} V_{\mu}^{a}(q, z)=0,
\end{equation}
where we have Fourier transformed with respect to the spacetime coordinates of
the boundary theory.
Splitting the vector field further as $V_{\mu}^{a}(q, z)=V(q,z)v^a_\mu$, the
on-shell action, given by the boundary term
\begin{equation} \label{equ:boundary_hw_model}
S=-\left.\frac{1}{2 g_{5}^{2}} \int d^{4} x \left( \frac{1}{z} V_{\mu}^{a} \partial_{z} V^{\mu a} \right) \right|_{z=\epsilon},
\end{equation}
is interpreted as generating functional for QCD flavor currents.
In both HW1 and HW2 models, boundary conditions on $V$ are such that there
is no contribution from $z=\zm$, while the conformal boundary at $z=0$ needs
regularization by a finite cutoff $z=\epsilon$ when imposing
$V(q,\epsilon)=1$.

Thus we find
\begin{equation} \label{equ:formula_HVP_HW_model}
\Pi_V\left(-q^{2}\right)=-\left.\frac{1}{g_{5}^{2} q^{2}} \frac{\partial_{z} V(q, z)}{z}\right|_{z=\epsilon}.
\end{equation}
The equations of motion \eqref{equ:EOM_hw_model} can be solved in terms of Bessel functions. Using the boundary conditions $V(q, \epsilon)=1$ and $\left. \partial_z V(q, z) \right|_{z=\zm} =0$ yields
\begin{equation}
V(q, z)=\left. \frac{z J_1(q z) Y_0\left(q \zm\right)-z Y_1(q z) J_0\left(q \zm\right)}{\epsilon  J_1(q \epsilon ) Y_0\left(q \zm\right)-\epsilon  Y_1(q \epsilon ) J_0\left(q \zm\right)} \right|_{\epsilon \rightarrow 0}.
\end{equation}
Plugging this result into equation \eqref{equ:formula_HVP_HW_model} yields
\begin{equation}
\Pi_{V}\left(-q^{2}\right)=-\frac{1}{g_{5}^{2}} \frac{1}{q \epsilon} \frac{J_{0}\left(q \zm \right) Y_{0}(q \epsilon)-Y_{0}\left(q \zm \right) J_{0}(q \epsilon)}{J_{0}\left(q \zm \right) Y_{1}(q \epsilon)-Y_{0}\left(q \zm \right) J_{1}(q \epsilon)},
\end{equation}
and expanding this at $\epsilon \rightarrow 0$ gives
\begin{equation} \label{equ:HVP_HW_small_epsilon}
\Pi_{V}\left(-q^{2}\right)=\frac{1}{g_{5}^{2}}\left[-\frac{\pi}{2} \frac{Y_{0}\left(q \zm \right)}{J_{0}\left(q \zm \right)}+\gamma-\log 2+\log q \epsilon+\mathcal{O}\left(\epsilon^{2}\right)\right],
\end{equation}
where $\gamma$ is the Euler-Mascheroni constant.
This expression is divergent for $\epsilon \rightarrow 0$ and has to be renormalized. Adding a counterterm
\begin{equation}
S_{\mathrm{c.t.}}(\mu)=\int d^{4} x\left(\frac{1}{2 g_{5}^{2}} \ln \epsilon \mu\right) \operatorname{tr}\left[F_{\mu \nu}(x, \epsilon)\right]^{2}
\end{equation}
to the action gives rise to an additional term
\begin{equation}
    \Pi_{V}^\mathrm{c.t.}=-\frac{1}{g_{5}^{2}} \log (\mu \epsilon),
\end{equation}
which cancels the divergent part of \eqref{equ:HVP_HW_small_epsilon}. The resulting renormalized vacuum polarization is 
\begin{equation}\label{HWPiVtimelike}
    \Pi_{V}^{\mathrm{ren}}\left(-q^{2}\right) = \lim _{\epsilon \rightarrow 0}\left[\Pi_{V}\left(-q^{2}\right)+\Pi_{V}^\mathrm{c.t.}\left(-q^{2}\right)\right]=\frac{1}{g_{5}^{2}}\left[-\frac{\pi}{2} \frac{Y_{0}\left(q \zm \right)}{J_{0}\left(q \zm \right)}+\gamma+\ln \frac{q}{2\mu}\right],
\end{equation}
where $\mu$ can be chosen such that $\Pi_{V}^{\mathrm{ren}}(0)=0$ holds, as
required for $\alpha$ in (\ref{equ:amu_hvp}) to be identified with the standard fine structure constant in the Thomson limit. For Euclidean momenta this yields
\be\label{HWPiVEucl}
    \Pi_{V}^{\mathrm{ren}}\left(Q^{2}\right) = 
    \frac{1}{g_{5}^{2}}\left[\frac{K_{0}\left(Q \zm \right)}{I_{0}\left(Q \zm \right)}+\ln\frac{Q\zm}{2}+\gamma\right].
\ee

As can be seen from (\ref{HWPiVtimelike}), in the time-like domain
the vacuum polarization function
has an infinite series of poles at $q^2=m_{n}^2$ with $m_n$ given
determined by $J_0(m_n\zm)=0$, corresponding to an infinite tower
of (stable) vector mesons (as expected in a large-$N_c$ limit).
The latter are described by normalizable solutions of
\begin{equation}
    \partial_{z}\left(\frac{1}{z} \partial_{z} \psi_{n}\right)+\frac{m_{n}^{2}}{z} \psi_{n}=0,
\end{equation}
with boundary conditions $\psi_n'(\zm)=0$, $\psi_n(0)=0$,
and are explicitly given by
\begin{equation}
    \psi_{{n}}(z)=\frac{\sqrt{2} z J_{1}\left(m_{n} z \right)}{\zm  J_{1}\left(m_{n} \zm  \right)}.
\end{equation}
The unrenormalized vector current correlator can then be represented as
\be\label{PiVsum}
    \Pi_V(q^2)=\sum_{n=1}^\infty \frac{F_n^2}{(q^2-m_n^2)m_n^2}
\ee
with decay constants $F_n$, defined as $\langle 0 |J_V^{a\mu}(0)|V_n^b \rangle=F_n \delta^{ab}\varepsilon^\mu$, given by
\be
    F_{n}=\lim_{\epsilon \rightarrow 0} \frac{1}{g_5}\psi_{n}^{\prime}(\epsilon) / \epsilon
    =\frac{1}{g_5}\frac{\sqrt{2} m_n}{\zm J_1(m_n \zm)}.
\ee

\subsubsection{Parameters of the HW1 model}

In the chiral limit, the HW1 model has only three free parameters,
$g_5$, $\zm$, and the chiral condensate described by $X(\zm)$.
In the application to HLbL contributions \cite{Leutgeb:2019gbz},
those were matched to the pion decay constant $f_\pi$, the $\rho$ meson mass,
and
$g_5$ was set such that the short-distance constraint on $\Pi_V$ from QCD
\cite{Shifman:1978bx,Reinders:1984sr}
\begin{equation} \label{equ:OPE_Pi_V}
    \Pi_{V}\left(Q^{2}\right)=\frac{N_{c}}{24 \pi^{2}}\left(1+\frac{\alpha_{s}}{\pi}\right) \log \left( \frac{Q^{2}}{\mu^{2}} \right)-\frac{\alpha_{s}}{24 \pi} \frac{N_{c}}{3} \frac{\left\langle G^{2}\right\rangle}{Q^{4}}+\frac{14 N_{c}}{27} \frac{\pi \alpha_{s}\langle q \bar{q}\rangle^{2}}{Q^{6}}
\end{equation}
is satisfied to leading order.\footnote{See \cite{Kurachi:2013cha} for a modification
of the HW1 model
which aims to incorporate the effects from the gluon condensate $\left\langle G^{2}\right\rangle$.}
This determines
\begin{equation}\label{eq:g5}
    \frac{1}{g_{5}^{2}}=\frac{N_{c}}{12 \pi^{2}},
\end{equation}
which holds true in other bottom-up models with bulk geometry that is
at least asymptotically AdS.

The HW cutoff $\zm$ directly determines the mass of the lightest vector meson,
which we choose as $m_\rho=775$ MeV, corresponding to
\begin{equation}
    \zm =3.103 \: \mathrm{GeV}^{-1}.
\end{equation}
This remains unchanged when finite quark masses are introduced in the HW1
model; the latter modify, however, axial-vector meson masses and
vector meson masses with open flavor quantum numbers \cite{Abidin:2009aj}.

\subsubsection{Parameters in the HW2 model}

In the inherently chiral
HW2 model due to Hirn and Sanz \cite{Hirn:2005nr}, chiral symmetry breaking
is implemented without a symmetry breaking bifundamental scalar $X$, through
different boundary conditions for vector and axial-vector fields. 

When $g_5$ and $z_0$ are chosen as above, the pion decay constant can no longer
be fitted to phenomenological values. In the application to HLbL contributions
for the muon anomalous magnetic moment \cite{Cappiello:2010uy,Leutgeb:2019zpq,Leutgeb:2019gbz,Cappiello:2019hwh}, which are dominated by the
coupling of pions to two photons, $f_\pi$ is fixed first, leaving the
choice of matching either the infrared parameter $m_\rho$ through $\zm$ or 
the ultraviolet behavior through $g_5$.
In \cite{Leutgeb:2019zpq,Leutgeb:2019gbz}, the former option, with physical
values for $f_\pi$ and $m_\rho$, is referred to as HW2 model. This
matches the short-distance constraints on transition form factors
and also the Melnikov-Vainshtein short-distant constraint \cite{Melnikov:2003xd} 
only at the level of 61\%. Conversely, the leading term in (\ref{equ:OPE_Pi_V})
is too large at the level of 164\%.
Matching
these constraints at the expense of a much too heavy rho meson (987 MeV) was called HW2(UV-fit).\footnote{%
These two choices correspond essentially to the parameters in ``Set 1'' and ``Set 2'' of
\cite{Cappiello:2019hwh}, where only the Hirn-Sanz model was considered.}

The symmetry breaking boundary conditions can in fact also be used in conjunction
with the symmetry breaking scalar $X$. This possibility has been proposed in
\cite{Domenech:2010aq} and also explored in \cite{Leutgeb:2021mpu} for
HLbL contributions to $a_\mu$, where it was referred to as HW3. 
In the vector sector, however, this coincides with the HW1
model so that our HVP results for the latter also pertain to the HW3 case.

Summarizing the values of the parameters for the three different fits we have, for $N_c=3$,
\begin{equation}
    \begin{aligned}
        \text { HW1 } :& \quad g_{5}=2 \pi, \quad  \zm =3.103 \: \mathrm{GeV}^{-1}; \\
        \text { HW2 } :& \quad g_{5}=4.932, \quad  \zm =3.103 \: \mathrm{GeV}^{-1}; \\
        \text { HW2 (UV-fit)}:& \quad g_{5}=2 \pi, \quad \zm =2.4359 \: \mathrm{GeV}^{-1}.
    \end{aligned}
\end{equation}

\subsection{Soft-wall model (SW)} \label{sec:softwall_model}

A shortcoming of the HW models is that the masses of highly excited vector mesons do
not rise as $m_n^2 \sim \sigma n$ as expected from linear confinement with string
constant $\sigma$, but instead like $m_n^2 \sim n^2$.

The so-called soft-wall model introduced in \cite{Karch:2006pv} (a precursor of which
appeared in \cite{Ghoroku:2005vt}) achieves
a strictly linear dependence of $m_n^2$ on $n$ by introducing a dilaton
$\Phi(z)=\kappa^2 z^2$ as an additional background field in the five-dimensional
Lagrangian
\cite{Karch:2006pv,Kwee:2007dd}
\begin{equation}
    S=\int d^{5} x e^{-\Phi(z)} \sqrt{g} \operatorname{Tr}\left\{|D X|^{2}+3|X|^{2}-\frac{1}{2 g_{5}^{2}}\left(F_{L}^{2}+F_{R}^{2}\right)\right\},
\end{equation}
while keeping the AdS metric (\ref{ds2AdS}).
The $z$-coordinate is, however, now unbounded from above, $z \in [0,\infty)$.

Solving the correspondingly modified field equation
\begin{equation} \label{equ:dgl_vector_mesons}
    \partial_{z}\left(\frac{e^{-\Phi}}{z} \partial_{z} V_{\mu}^{a}\right)+\frac{q^{2} e^{-\Phi}}{z} V_{\mu}^{a}=0
\end{equation}
with the boundary conditions $V(q, \epsilon)=1$ in the limit $\epsilon \rightarrow 0$ and $\lim_{z \rightarrow \infty} V(q, z)=0$ gives \cite{Grigoryan:2007my}
\begin{equation}
    V(q, z)=\Gamma\left(1-\frac{q^{2}}{4 \kappa^{2}}\right) U\left(\frac{-q^{2}}{4 \kappa^{2}}, 0,(\kappa z)^{2}\right),
\end{equation}
with the confluent hypergeometric function of second kind $U$ (also known as Tricomi function). Plugging this into 
\begin{equation} \label{equ:self_energy_sw}
\Pi_{\mathrm{V}}\left(-q^{2}\right)=-\left.\frac{e^{-\Phi}}{g_{5}^{2}} \frac{\partial_{z} V(q, z)}{q^{2} z}\right|_{z=\epsilon} = -\left.\frac{1}{g_{5}^{2}} \frac{\partial_{z} V(q, z)}{q^{2} z}\right|_{z=\epsilon},
\end{equation}
switching to the Euclidean momentum $Q^2=-q^2$ and expanding in a series for small $\epsilon$ gives
\begin{equation}
    \Pi_{V}\left(Q^{2}\right) = \frac{1}{2 g_{5}^{2}} \left[\psi\left(\frac{Q^2}{4 \kappa^2}+1\right)+\ln \left(\kappa^2\epsilon^2\right)+2 \gamma \right] + \mathcal{O}(\epsilon),
\end{equation}
where $\psi(z)=d\ln\Gamma(z)/dz.$
Renormalizing as above and using $\psi(0)=-\gamma$ we find
\be
    \Pi^\mathrm{ren}(Q^2)=\frac{1}{2 g_{5}^{2}} \left[\psi\left(\frac{Q^2}{4 \kappa^2}+1\right)+\gamma \right],
\ee
with $g_5$ fixed as in the HW1 model.

From the poles of the digamma function $\psi$ one can see that the spectrum
of vector mesons is now given by
\be
    m_n^2=4\kappa^2 n, \qquad n=1,2,3,\ldots.
\ee
Alternatively, the hadronic vacuum polarization can be calculated as above
from the normalizable solutions given by \cite{Karch:2006pv,Grigoryan:2007my}
\be
\psi_n(z)=\sqrt{\frac{2}{n}}\kappa^2 z^2 L_{n-1}^1(\kappa^2 z^2),
\ee
where the $L_n^1$ are the generalized Laguerre polynomials of order one, 
leading to decay constants
\begin{equation}
    F_{n}= \frac{1}{g_5}\psi_n''(0) =\frac{1}{g_5} \kappa^2 \sqrt{8 n}.
\end{equation}

\subsection{Interpolating models}

The phenomenological study of pion form factors for HW and SW models in \cite{Kwee:2007dd} 
came to the conclusion that the HW1 model generally performed better than the SW model.
In \cite{Kwee:2007dd,Kwee:2007nq}, a simple interpolating model was proposed that combines
features of HW and SW models, whereas in \cite{Casero:2007ae, Iatrakis:2010zf, Iatrakis:2010jb} a more sophisticated version including a dynamical tachyon
for chiral symmetry breaking was developed, which achieves a similar behavior.

\subsubsection{Semi-hard wall model (SHW)}

In the model proposed by Kwee and Lebed in \cite{Kwee:2007dd,Kwee:2007nq}, a
semi-hard wall was set up by replacing the dilaton background of the SW model $e^{-\Phi}=e^{-\kappa^2 z^2}$ with a background given by
\begin{equation} \label{equ:dilaton_background_interpol}
    e^{-\Phi(z)}=\frac{e^{\lambda^{2} \zm ^{2}}-1}{e^{\lambda^{2} \zm ^{2}}+e^{\lambda^{2} z^{2}}-2}.
\end{equation}
The HW model is recovered by $\lambda \zm\to\infty$ at fixed $\zm$,
whereas at large $z$ the dilaton behaves as $\Phi(z)\sim \lambda^2 z^2$.

In \cite{Kwee:2007nq} two sets of parameters were considered,
involving $\lambda \zm=2.1$ and $1$,
where the first choice was found to give a good agreement with the pion form factor $F_{\pi}(Q^{2})$ comparable to that of the HW1 model, albeit at the cost of not matching the pion decay constant very well; the second choice led to a similar prediction for $F_{\pi}(Q^{2})$ as the SW model.

In the following we shall choose the two free parameters $\lambda$ and $\zm$ such
that in addition to the mass of the lightest rho meson this model
reproduces the mass of $\rho(1450)$, which \cite{PDG20} lists as 1465$\pm25$ MeV.
This leads to
\be\label{SHWparms}
     \lambda \zm =1.697, \quad \zm =2.9738\, \text{GeV}^{-1},
\ee
which is right in between the two parameter sets explored in \cite{Kwee:2007nq}.

With (\ref{equ:dilaton_background_interpol}) closed analytical results are
no longer available and one has to resort to numerical solutions of the
equations of motion for the vector modes. 
In contrast to the SW model, $m_n^2$ becomes a linear function of $n$ only for large $n$,
where the spacing is determined by $m_{n+1}^2-m_n^2\sim 4\lambda^2 n$ ($\approx 1.3 \,\text{GeV}^2$
for our choice (\ref{SHWparms})).
The decay constants are again given by $F_n=\psi''(0)/g_5$ with $g_5$ still determined by (\ref{eq:g5}).
Table \ref{tab:meson_masses_decayconst} lists the results for the first 8 modes.

The full subtracted self energy function with $\Pi_{V}^{\text{ren}}(0)=0$ can then be calculated through 
\begin{equation} \label{equ:kir_PI_ren}
    \Pi_{V}^{\text {ren }}\left(-q^{2}\right) =\sum_{n=1}^{m} \frac{q^{2} F_{{n}}^{2}}{\left(q^{2}-m_{{n}}^{2}\right) m_{{n}}^{4}}+\mathcal{O}\left(\frac{q^{2}}{m_{{m+1}}^{2}} \right).
\end{equation}

\subsubsection{Tachyon condensation model (TC)} \label{sec:tachyon_model_Kiritsis}

Finally we consider a more sophisticated but still relatively simple bottom-up model, developed in \cite{Casero:2007ae, Iatrakis:2010zf, Iatrakis:2010jb}, where chiral symmetry breaking is implemented
by a brane-antibrane effective action with an open string tachyon mode as
proposed by Sen \cite{Sen:2003tm}.
This is based on a pair of Dirac-Born-Infeld type Lagrangians augmented by a tachyon
potential $V(|T|)$, which in the case of a single flavor reads
\begin{equation} \label{equ:action_mesons_Kiritsis}
    S=\int d^{4} x d z V(|T|)\left(\sqrt{\operatorname{det} \mathbf{A}^{(L)}}+\sqrt{\operatorname{det} \mathbf{A}^{(R)}}\right),
\end{equation}
where
\begin{equation}
    \mathbf{A}^{(L,R)}_{M N}=g_{M N}+\frac{2 \pi \alpha^{\prime}}{g_{V}^{2}} F_{M N}^{(L,R)}+\pi \alpha^{\prime} \lambda\left[\left(D_{M} T\right)^{*}\left(D_{N} T\right)+\left(D_{N} T\right)^{*}\left(D_{M} T\right)\right],
\end{equation}
and
\begin{equation}
    D_{M} T=\left(\partial_{M}+i A_{M}^{(L)}-i A_{M}^{(R)}\right) T, \quad T=\tau e^{i \theta},
\end{equation}
with a Gaussian potential
\begin{equation}
    V=\mathcal{K} e^{-\frac{1}{2} \mu^{2} \tau^{2}}.
\end{equation}
The background geometry is derived from a six-dimensional
AdS soliton, which has been proposed by Kuperstein and Sonnenschein as
a holographic model of four-dimensional Yang-Mills theory \cite{Kuperstein:2004yf}.
It is given by
\begin{equation} \label{equ:metric_kiritsis_induced}
    d s_{5}^{2} = g_{t t} d t^{2} - g_{z z} d z^{2} - g_{x x} d x_{3}^{2}=\frac{R^{2}}{z^{2}}\left[d x_{1,3}^{2} - f_{\Lambda}^{-1} d z^{2}\right],
\end{equation}
where $f_{\Lambda}=1-(z/z_{\Lambda})^{5}$. 

Similarly to the case of the bifundamental scalar $X$ in the HW and SW models,
the scalar $T$ mediates chiral symmetry breaking by its possible vacuum solutions.
In order to match the scaling dimension of a quark bilinear, the AdS/CFT
correspondence requires to set the mass of the field $\tau$ to
\be
m_\tau^2 R^2=-\frac{R^2\mu^2}{2\pi\alpha'\lambda}=-3,
\ee
which leads to the differential equation for its profile $\tau(z)$
\begin{equation} \label{equ:dgl_tachyon_kiritsis}
    \tau^{\prime \prime}-\frac{4 \mu^{2} z f_{\Lambda}}{3} \tau^{\prime 3}+\left(-\frac{3}{z}+\frac{f_{\Lambda}^{\prime}}{2 f_{\Lambda}}\right) \tau^{\prime}+\left(\frac{3}{z^{2} f_{\Lambda}}+\mu^{2} \tau^{\prime 2}\right) \tau=0
\end{equation}
and a UV asymptotic behavior parametrized by two constants $c_1$ and $c_3$
\begin{equation} \label{equ:UV_tachyon}
    \tau=c_{1} z+\frac{\mu^{2}}{6} c_{1}^{3} z^{3} \log z+c_{3} z^{3}+\mathcal{O}\left(z^{5}\right).
\end{equation}
Choosing the source parameter $c_1$ corresponding to the quark mass, the parameter $c_3$
is tuned such that the tachyon diverges exactly at $z=z_\Lambda$.

The parameter $\mu$ in the tachyon potential does not have a physical meaning as it can be absorbed in the definition of $\tau$; in the following it will be set to $\mu^2=\pi$.
In \cite{Iatrakis:2010zf} 
a fit of light unflavored mesons (composed of $u$ and $d$ quarks) gave
\begin{equation} \label{equ:values_tachyon_model}
    z_{\Lambda}^{-1}=522 \mathrm{MeV}, \quad \mathrm{c}_{1}=0.0125 z_{\Lambda}^{-1}.
\end{equation}
We will refer to this as TC (fit 1). In \cite{Iatrakis:2010jb} the parameters are chosen as
\begin{equation} \label{equ:values_tachyon_model_2}
    z_{\Lambda}^{-1}=549 \mathrm{MeV}, \quad \mathrm{c}_{1}=0.0094 z_{\Lambda}^{-1},
\end{equation}
which gives a slightly higher mass for the lightest rho meson [referred
to as TC (fit 2) in the following]. 

\begin{figure}
    \centering
    \includegraphics[width=0.55\textwidth]{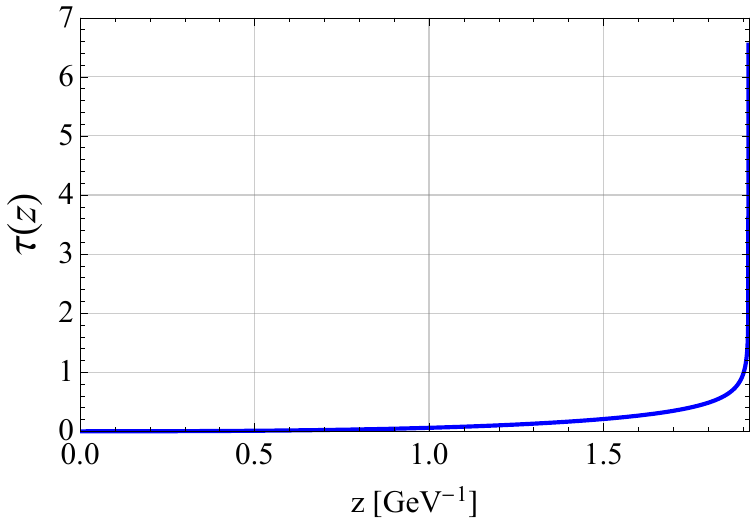}
    \caption[Plot of the tachyon field $\tau(z)$]{The 
    tachyon field $\tau(z)$ for the parameters in \eqref{equ:values_tachyon_model}. 
    }
    \label{fig:tachyon_tau}
\end{figure}

The function $\tau(z)$ we obtain as the solution of equation \eqref{equ:dgl_tachyon_kiritsis} is plotted in figure \ref{fig:tachyon_tau}.
The value we obtain for the other constant is
\begin{equation}
c_3 \approx 0.37 z_{\Lambda}^{-3}.
\end{equation}

Given $\tau(z)$, one can proceed by expanding the action \eqref{equ:action_mesons_Kiritsis} up to quadratic order in the fields as in the other models above.

For the vector gauge fields
$V_{\mu}=[A_{\mu}^{(L)}+A_{\mu}^{(R)}]/{2}$ and corresponding field strength $V_{\mu\nu}$
the action
up to quadratic order reads
\begin{equation} \label{equ:action_quadr_meson_kiritsis}
    S_{V} = \frac{\left(2 \pi \alpha^{\prime}\right)^{2}}{g_{V}^{4}} \mathcal{K} \int d^{4} x d z e^{-\frac{1}{2} \mu^{2} \tau^{2}}\left[\frac{1}{2} \tilde{g}_{z z}^{\frac{1}{2}} V_{\mu \nu} V^{\mu \nu}+g_{x x} \tilde{g}_{z z}^{-\frac{1}{2}} \partial_{z} V_{\mu} \partial_{z} V^{\mu}\right],
\end{equation}
where we have defined $\tilde{g}_{z z}=g_{z z}+2 \pi \alpha^{\prime} \lambda\left(\partial_{z}\langle\tau\rangle\right)^{2}$.
This leads to the mode equations
\begin{equation} \label{equ:dgl_eigens_Kiritsis}
    -\frac{1}{e^{-\frac{1}{2} \mu^{2} \tau^{2}} \tilde{g}_{z z}^{\frac{1}{2}}} \partial_{z}\left(e^{-\frac{1}{2} \mu^{2} \tau^{2}} g_{x x} \tilde{g}_{z z}^{-\frac{1}{2}} \partial_{z} \psi_{n}(z)\right)=m_{n}^{2} \psi_{n}(z).
\end{equation}

As in the SHW model, solutions have to be obtained by relying on numerical
calculations, which is best done by transforming the differential equations
to Liouville normal form (see the Appendix of \cite{Iatrakis:2010jb}).
The vector correlator,
which is given by
\begin{equation}
    \Pi_{{V}}\left(-q^{2}\right)= - \left. 2 \frac{\left(2 \pi \alpha^{\prime}\right)^{2} \mathcal{K} R}{g_{V}^{4}} \frac{\partial_{z} V(q, z)}{q^{2} z}\right|_{z=\epsilon}.
\end{equation}
for a solution with boundary condition $V(q,0)=1$,
can then be calculated alternatively by summing a sufficiently
large number of modes according to (\ref{equ:kir_PI_ren}). The behavior
at large $q^2$ can be shown \cite{Iatrakis:2010jb} to be of a form similar to the SW model.
Matching to the leading-order term of the OPE result leads to
\begin{equation}
    \frac{2\left(2 \pi \alpha^{\prime}\right)^{2} \mathcal{K} R}{g_{V}^{4}} = \frac{N_c}{12 \pi^2}.
\end{equation}
Like the SHW model, the TC model has a linear dependence of $m_n^2$ on $n$ only for large $n$,
where it can be shown \cite{Iatrakis:2010jb} that the spacing is given by 
$m_{n+1}^2-m_n^2\sim 6 z_{\Lambda}^{-2}$ ($\approx 1.6 \,\text{GeV}^2$
and $ 1.8 \,\text{GeV}^2$ for (\ref{equ:values_tachyon_model}) and (\ref{equ:values_tachyon_model_2}), respectively).

\section{Numerical results}\label{sec:numerics}

\subsection{Masses and decay constants}

In Table \ref{tab:meson_masses_decayconst} we list our results for the masses
and decay constants
of the vector mesons in the various models.
In the HW1, SW, SHW models we have fixed $m_1=775$ MeV while setting $g_5$ such
that the asymptotic behavior of the vector correlator is matched.
The extra parameter
in the SHW model was used to additionally match $m_2=1465$ MeV, while in
the TC model we have considered the two sets presented in \cite{Iatrakis:2010zf}
and \cite{Iatrakis:2010jb}. The simpler HW2 model, which instead has fewer
parameters, is considered in the two versions
used in \cite{Leutgeb:2019gbz,Cappiello:2019hwh} for evaluating the
HLbL contribution to the muon $g-2$: an IR fit where $m_\rho$ and $f_\pi$ is matched
but short-distance constraints are only satisfied at the level of 61\%,
and a UV-fit where $f_\pi$ and the short-distance behavior are correct but
$m_\rho$ too heavy by 27\%.

\begin{figure}[htbp] 
    {\centering
    \includegraphics[width=0.58\textwidth]{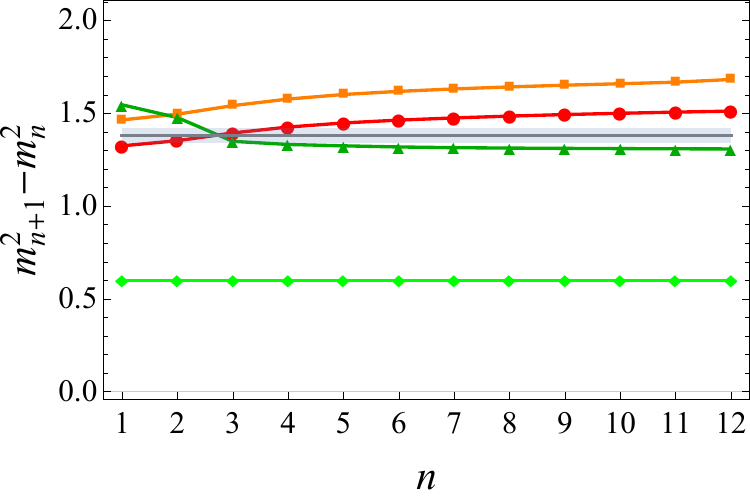} 
    }
    \caption{Plots of $m_{n+1}^2-m_n^2$ in GeV$^2$ in models where $m_n^2\sim n$ for large $n$: the SW model (light green), the SHW model (darker green), and the TC model (fit $1$ in red and fit $2$ in orange). The gray line is the phenomenological value given in \cite{Masjuan:2012gc} as $1.38(4)\,\text{GeV}^2$, with the gray shaded region representing its uncertainty.}
    \label{fig:masses}
\end{figure}

In the HW models, the masses of excited rho mesons rise very quickly,
asymptotically like $m_n^2\sim n^2$, whereas the SW, SHW, and TC models
have $m_n^2\sim n$ as required by linear confinement. 
While in the HW models the first excited rho meson has a mass significantly
higher than the experimental value $m_2=1465$ MeV,
in the simple SW this value is 25\% too low, this can be remedied in the SHW by
our choice for its parameters,
and also by the overall fits \cite{Iatrakis:2010zf,Iatrakis:2010jb} in the TC model.
In the SHW and TC models, also $m_3$ and $m_4$ are well compatible with
the next states on the radial Regge trajectory, which in \cite{Masjuan:2012gc}
were assumed to be $\rho(1900)$ and $\rho(2150)$.
In Fig.~\ref{fig:masses} we plot the increments of the masses squared for
the first 12 modes in the models with an asymptotically linear behavior, 
which shows that the simple SW model has a much
smaller value ($m_{n+1}^2-m_n^2\equiv m_\rho^2 \approx 0.601\,\text{GeV}^2$) 
than the SHW and TC models. The latter are in fact
closer to the observed slope of radial Regge trajectories, which
in \cite{Masjuan:2012gc} was determined as $1.38(4)\,\text{GeV}^2$.

Regarding decay constants, sufficient experimental information is only available for the
lightest rho meson 
through $\Gamma(\rho^0\to e^+e^-)=7.04(6)$ keV \cite{PDG20}, which yields \cite{Donoghue:1992dd}
\be\label{Frho-exp}
F_{\rho^0}^2=3m_\rho^3\Gamma(\rho^0\to e^+e^-)/(4\pi\alpha^2)=[348(1)\,\text{MeV}]^4.
\ee
The largest deviations from this result, of about 25\% and 20\%, respectively, 
are found in the SW model and in
the HW2(UV-fit) case, whereas the HW1 model are merely 5\% too low.
Note that $F_n\propto g_5^{-2}$, which implies that one could match the experimental result
by a corresponding adjustment of $g_5$.

\begin{table}[t]
    \begin{tabular}{c|cc|cc|cc|cc|cc}
    \toprule
     & \multicolumn{2}{c}{HW1} & \multicolumn{2}{c}{HW2(IR$|$UV-fit)} & \multicolumn{2}{c}{SW} & \multicolumn{2}{c}{SHW}  & \multicolumn{2}{c}{TC(fit$1|2$)}   \\
    ${n}$ & $m_{n}$ & $F_{n}^{1/2}$ & $m_{n}$ & $F_{n}^{1/2}$ & $m_{n}$ & $F_{n}^{1/2}$ & $m_{n}$ & $F_{{n}}^{1/2}$ & $m_{n}$ & $F_{{n}}^{1/2}$  \\
    \colrule
    $1$ & \textit{775} & $329.1$ & \textit{775}$|987.2$ & 
    $371.4|419.2$ & \textit{775} & $260.0$ &  \textit{775} & 314.0  & $765.4|803.9$ & $313.2|329.0$ \\
    $2$ & $1779$ & $615.8$ & $1779|2266$ & 
    $695.1|784.5$ & $1096$ & $309.2$ & \textit{1465}  &  458.5 & $1382|1453$ & $418.0|438.9$ \\
    $3$ & $2789$ & $863.3$ & $2789|3553$ & 
    $974.4|1100$ & $1342$ & $342.2$ & 1903  &  498.7 & $1806|1899$ & $488.7|513.0$ \\
    $4$ & $3800$ & $1089$ & $3800|4841$ & 
    $1229|1387$ & $1550$ & $367.7$ & 2230  & 540.0  & $2158|2269$ & $538.4|564.9$ \\
    $5$ & $4812$ & $1300$ & $4812|6130$ & 
    $1467|1656$ & $1733$ & $388.8$ & 2511  & 570.7  & $2466|2593$ & $577.3|605.4$ \\
    $6$ & $5824$ & $1500$ & $5824|7419$ & 
    $1693|1911$ & $1898$ & $406.9$ & 2762  & 597.4  & $2744|2885$ & $610.4|639.6$ \\
    $7$ & $6836$ & $1692$ & $6836|8708$ & 
    $1909|2155$ & $2050$ & $422.9$ & 2991  & 621.4  & $2999|3153$ & $639.4|669.7$ \\
    $8$ & $7848$ & $1876$ & $7848|9997$ & 
    $2118|2390$ & $2192$ & $437.2$ & 3203  & 642.8  & $3236|3402$ & $665.5|696.6$ \\
    \botrule
    \end{tabular}
    \caption[Vector meson masses $m_{n}$ and decay constants $F_{n}^{1/2}$ of the HW models]{Vector meson masses $m_{n}$ and decay constants $F_{n}^{1/2}$ in $\mathrm{MeV}$.
    }
    \label{tab:meson_masses_decayconst}
\end{table}

\subsection{$N_f=2$ LO-HVP contribution to $a_\mu$}

In Table {\ref{tab:amu_results} we finally give the results for the leading-order
HVP contribution to the anomalous magnetic moment of the muon with two light flavors, $a_{\mu(N_f=2)}^\textrm{LO-HVP}$,
by using equation \eqref{equ:amu_hvp}.
As mentioned above, for the HW and SW models we can use
closed form expressions for $\Pi_V$.\footnote{In \cite{Hong:2009jv} $\Pi_V$ in the HW1 model was
calculated by truncating the infinite sum in (\ref{PiVsum}) at $n=4$, which produces a result that
is about 1\% lower than the full contribution. With the slightly different choice of $m_1=775.49$ MeV of  \cite{Hong:2009jv}, 
we would obtain $a_{\mu (N_f=2)}^{\text{LO-HVP}}=476.4\times 10^{-10}$, while
the truncated result given in \cite{Hong:2009jv} is $a_{\mu (N_f=2)}^{\text{LO-HVP}}=470.5\times 10^{-10}$.}
For the SHW and TC models we rely on numerical results and the expansion \eqref{equ:kir_PI_ren}. 
The corresponding integrands in the master formula (\ref{equ:amu_hvp}) are displayed in Fig.\ \ref{fig:integrand}.

\begin{figure}
    \centering
    \includegraphics[width=0.6\textwidth]{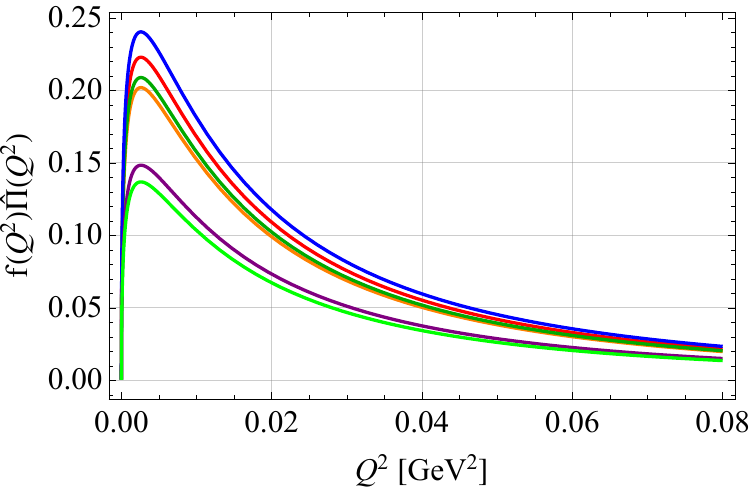}
    \caption[]{Plot of the integrand $f(Q^2) \hat{\Pi}(Q^2)=f(Q^{2}) 4\pi^2 \Pi_{\text{em}}^{\mathrm{had}}(Q^{2})$ in \eqref{equ:amu_hvp} for the HW1 model (blue), SW model (light green), SHW model (darker green), and the TC model (red for fit $1$ and orange for fit $2$). The HW2 model is only shown in the version UV-fit (purple), where the
    UV behavior is reproduced at the expense of an unrealistically heavy rho meson; when $m_\rho$ and $f_\pi$ are fitted, the integrand peaks at about 0.39 and the UV behavior is about a factor 1.6 too large.}
    \label{fig:integrand}
\end{figure}

\begin{table}
    \centering
    \begin{tabular}{lcc}
    \toprule\\[-12pt]
     & $a_{\mu (N_f=2)}^{\text{LO-HVP}}$ & mismatch\\[4pt]
    \colrule
    HW1          & 476.9 & 0.86\\ 
    HW2(IR$|$UV-fit)         & 773.9$|$304.0 & 1.39$|$0.55\\ 
    SW & $276.4$   & 0.50\\ 
    SHW  & $415.4$ &0.75\\
    TC (fit 1$|$2) & $442.3|403.6$  & 0.79$|$0.72 \\ \botrule
    \end{tabular}
    \caption{Values of $a_{\mu}^{\text{LO-HVP}}$ with $N_f=2$ for the different hQCD models in multiples of $10^{-10}$. The last column gives the ratio of these results over $a_{\mu(\pi\pi,\pi\pi\pi,\pi\gamma)}^\text{LO-HVP}$ as obtained in the dispersive approach (\ref{amudisp-rho+omega}).}
    \label{tab:amu_results}
\end{table}

\begin{table}[t]
    \centering
    \begin{tabular}{lcc}
    \toprule
     &  DHMZ19 \cite{Davier:2019can} &  KNT19 \cite{Keshavarzi:2019abf} \\
    \colrule 
    $\pi^{+} \pi^{-}$ & $508(3)$ & $504(2)$ \\
    $\pi^{+} \pi^{-} \pi^{0}$ & $46(1)$& $47(1)$ \\
    $\pi^0 \gamma$ & 4.4(1) & 4.6(1) \\
    \colrule
    Sum & $558(4)$ & $555(3)$ \\
    \botrule
    \end{tabular}
    \caption[Contributions of the different decay channels to $a_{\mu}^{\text{LO-HVP}}$ in the dispersive approach]{Rounded contributions of decay channels with one, two and three pions to $a_{\mu}^{\text{LO-HVP}}$ in multiples of $10^{-10}$ \cite{Aoyama:2020ynm}.}
    \label{tab:amu_disp1}
\end{table}

With the exception of the model HW2 (with fitted $m_\rho$ and $f_\pi$), 
where the asymptotic behavior of $\Pi_V$
is about a factor 1.6 larger than the OPE result, the holographic results
are much smaller than the results obtained for the $N_f=2$ contributions in
dispersive and lattice approaches (the latter are about \cite{Aoyama:2020ynm} $590\times10^{-10}$ and
\cite{Borsanyi:2020mff} $640\times10^{-10}$, respectively).
However, the holographic QCD models should be viewed as (more or less crude) large-$N_c$ approximations. HVP contributions with multi-hadron states such as four pions correspond
to higher-order contributions in the large-$N_c$ expansion so that it appears more reasonable
to compare only with contributions from
intermediate states corresponding to the $\rho$ and $\omega$ channels, which
are dominated by two and three-pion states as well as $\pi^0\gamma$.
Table \ref{tab:amu_disp1} lists their contributions
to $a_\mu$ according to \cite{Davier:2019can} and \cite{Keshavarzi:2019abf}, which
combined give approximately 
\be\label{amudisp-rho+omega}
a_{\mu(\pi\pi,\pi\pi\pi,\pi\gamma)}^\text{LO-HVP}=557(3)\times10^{-10}.
\ee

While the HW2 model in its two versions brackets this result, with rather
large deviations on either side, the HW1 model, where $m_\rho$ and $f_\pi$
as well as the short-distance behavior can be fitted simultaneously,
is only a factor 0.86 smaller, and thus comes closest of all models considered here.

The smallest result, at only 50\%, is
obtained with the SW model. As we have seen above, the SW model with a
strictly linear dependence of $m_n^2$ on $n$ underestimates the masses
of all excited rho mesons. While this should tend to overestimate $a_\mu^\text{LO-HVP}$, the decay constant squared of the ground-state rho meson is only at 30\% of its experimental
value, which is thus responsible for the strong attenuation.
However, the simple modification (\ref{equ:dilaton_background_interpol}) in the SHW model,
which leads to a much improved mass spectrum, also brings the decay constants closer
to realistic values, yielding a result for $a_\mu^\text{LO-HVP}$ that is
only 25\% below (\ref{amudisp-rho+omega}).
The more sophisticated TC model turns out to
be comparable, coming somewhat closer with parameters of fit 1.

Thus all models which reproduce $m_\rho$, $f_\pi$ and $F_{\rho^0}$ reasonably well
also do not deviate too strongly from the dispersive
result for $a_\mu^\text{LO-HVP}$, but uniformly underestimate it.
Since the latter is proportional to $g_5^{-2}$, this suggests that
its value, obtained from matching the leading-order term in the vector correlator (\ref{equ:OPE_Pi_V}),
should be corrected to account for the next-to-leading order term, which is indeed positive. 
Exactly such a correction was proposed by two of us
in the evaluation of the HLbL contribution within the (massive) HW1 and HW3 models
\cite{Leutgeb:2021mpu,Leutgeb:2021bpo}, where it has the effect of
reducing the holographic HLbL result, as this brings
the asymptotic behavior of transition form factors down by amounts that are roughly consistent with perturbative corrections to the leading-order pQCD results at moderately high $Q^2$ values \cite{Melic:2002ij,Bijnens:2021jqo}. At the same time, the coefficient
of the logarithm in the asymptotic expression (\ref{equ:OPE_Pi_V}) is increased by a similar amount, which is consistent with the next-to-leading order terms in this expression.

In the case of the HW1 model, $F_{\rho^0}$ can be matched by reducing $g_5^2$
by a factor $0.9$. This happens to bring the HW1 result for the $\pi^0$ pole
contribution to the HLbL part of $a_\mu$ into perfect agreement with
the dispersive result \cite{Leutgeb:2021mpu}: $a_{\mu \text{(HW1,3)}}^{\pi^0}=(6.17\ldots 6.39)\times 10^{-10}$ while \cite{Hoferichter:2018kwz} $a_{\mu \text{(disp.)}}^{\pi^0}=6.26^{+30}_{-25}\times 10^{-10}$.

With $F_{\rho^0}$ matched, 
the HW1 result for $a_\mu^\text{LO-HVP}$ becomes correspondingly
larger, namely $533.2\times 10^{-10}$, which is less than 5\% smaller than
the dispersive result (\ref{amudisp-rho+omega}).\footnote{Interestingly, in \cite{Kurachi:2013cha}
it has been argued that inclusion of the effects of a gluon condensate within
a modified HW1 model leads to an increase of about 6\% of the holographic
value for $a_\mu^\text{LO-HVP}$.}

\section{Conclusion}\label{sec:conclusion}

By considering a number of simple bottom-up holographic-QCD models we have
found that their quantitative predictions are too spread out to be of help
with the task of determining the HVP contribution to the anomalous magnetic
moment of the muon, which is currently afflicted by the largest uncertainty
with regard to the ongoing efforts of testing the Standard Model by
a new round of experiments. However, a comparison of the holographic
results for the LO-HVP contribution with the existing data-driven results
at or below percent accuracy allows us to assess the various holographic
models with regard to their ability to account for the relevant interactions
between hadrons and photons. This is useful because holographic QCD
can provide interesting estimates for HLbL contributions, where
conventional approaches have uncertainties that are comparable with or larger than
expected errors in the large-$N_c$ limit that holographic QCD is based upon.\footnote{For example, the contribution of axial vector mesons is
currently assigned a 100\% uncertainty in \cite{Aoyama:2020ynm}.}

In particular, we have considered the holographic SW and HW models that have been
used previously for estimating the HLbL contributions of pseudoscalars
and axial-vector mesons (see the recent review \cite{Leutgeb:2021bpo}),
and we have also explored two simple extensions that aim at interpolating
between the HW and SW models, while keeping their respective advantages.

We have found that the original HW1 model \cite{Erlich:2005qh} turned
out to come closest to the phenomenological value of the rho meson
decay constant as well as to the value for $a_{\mu (N_f=2)}^{\text{LO-HVP}}$
obtained in dispersive approaches. The somewhat simpler HW2 model, which was
used in holographic calculations of the axial-vector contribution
in two versions which either fit IR or UV constraints, brackets
the latter with rather large deviations in both directions.
The SW model turns out to give the worst fit, but already the simple
improvement of a semi-hard wall as proposed in \cite{Kwee:2007dd,Kwee:2007nq}
reduces the deviation considerably; the more sophisticated TC model
achieves roughly the same with the parameters considered previously
in \cite{Casero:2007ae, Iatrakis:2010zf, Iatrakis:2010jb}.

In the HW1 model, the LO-HVP result is simply proportional to the
coupling $g_5^{-2}$ determining the asymptotic behavior of the
vector correlator. Reducing $g_5^2$ by a factor 0.9 or 0.85 has been
proposed in \cite{Leutgeb:2021mpu,Leutgeb:2021bpo} as a simple
way to account for next-to-leading order QCD effects for the
large-$Q^2$ behavior of transition form factors. In the case of
the rho meson decay constant, a factor of 0.9 leads to a perfect
fit with the phenomenological value and a result for $a_{\mu (N_f=2)}^{\text{LO-HVP}}$
that is only 5\% too small. As shown already in \cite{Leutgeb:2021mpu,Leutgeb:2021bpo},
the same reduction of $g_5^2$ brings about a perfect agreement of the pion
pole contribution in the HW1 model with the data-driven result of
\cite{Hoferichter:2018kwz}. We interpret this as a support
for the predictions for pseudoscalar and axial-vector meson contributions
obtained by two of us in various versions of the HW1 and HW3 model \cite{Leutgeb:2021mpu,Leutgeb:2021bpo}, where a theoretical error
was formed by taking the unchanged results of these models as
upper bound and those with $g_5^2$ reduced by a factor of 0.85 as lower bound.
The corresponding values with the factor 0.9 could then be regarded
as the best guess within these models.\footnote{We do not reproduce these
numbers here. They can be easily obtained by applying the correction factors
given in Table III of \cite{Leutgeb:2021mpu} to the results in Table II therein.}

As an outlook we would like to refer to the many possible improvements
that can be considered for bottom-up holographic QCD models.
In \cite{Kurachi:2013cha} it has already been shown that
incorporating the effects of a gluon condensate within
a modified HW1 model leads to an increase of about 6\% of the holographic
value for $a_\mu^\text{LO-HVP}$, bringing it very close to the data-driven
result. It would be interesting to study even more extensions such
as models that relax the assumption $N_f\ll N_c$ of the 't Hooft limit
\cite{Jarvinen:2011qe}.


\acknowledgments

J.~L.\ was supported by the FWF doctoral program
Particles \& Interactions, project no. W1252-N27,
and FWF project no.\ P33655. This work has been partially funded by the Deutsche Forschungsgemeinschaft (DFG, German Research Foundation) under Germany’s Excellence Strategy - EXC-2094 - 390783311 and the DFG grant BSMEXPEDITION.


\bibliographystyle{JHEP}
\bibliography{hvp}

\end{document}